\documentclass[pdflatex,sn-nature]{sn-jnl}

\usepackage{graphicx}%
\usepackage{multirow}%
\usepackage{amsmath,amssymb,amsfonts}%
\usepackage{amsthm}%
\usepackage{mathrsfs}%
\usepackage[title]{appendix}%
\usepackage{xcolor}%
\usepackage{textcomp}%
\usepackage{manyfoot}%
\usepackage{booktabs}%
\usepackage{algorithm}%
\usepackage{algorithmicx}%
\usepackage{algpseudocode}%
\usepackage{listings}%

\theoremstyle{thmstyleone}%
\theoremstyle{thmstyletwo}%

\theoremstyle{thmstylethree}%

\begin{document}

\title[Demonstration of a planar multimodal periodic filter at THz frequencies]{Demonstration of a planar multimodal periodic filter at THz frequencies}


\author[1,2]{\fnm{Ali} \sur{Dehghanian}}\email{adehghanian@uvic.ca}\equalcont{These authors contributed equally to this work.}

\author[1,2]{\fnm{Mohsen} \sur{Haghighat}}\email{mohsenh@uvic.ca}

\author[1]{\fnm{Thomas} \sur{Darcie}}\email{tdarcie@uvic.ca}

\author*[1,2]{\fnm{Levi} \sur{Smith}}\email{levismith@uvic.ca}
\equalcont{These authors contributed equally to this work.}

\affil*[1]{\orgdiv{Department of Electrical and Computer Engineering}, \orgname{University of Victoria}, \orgaddress{\street{3800 Finnerty Rd}, \city{Victoria}, \postcode{V8P 5C2}, \state{BC}, \country{Canada}}}

\affil[2]{\orgdiv{Centre for Advanced Materials and Related Technology (CAMTEC)}, \orgname{University of Victoria}, \orgaddress{\street{3800 Finnerty Rd}, \city{Victoria}, \postcode{V8P 5C2}, \state{BC}, \country{Canada}}}


\abstract{This paper presents a planar multimodal periodic filter that is constructed from alternating sections of coplanar stripline and the odd-mode of a finite-ground plane coplanar waveguide on a 1 \textmu m silicon nitride substrate to facilitate operation at THz frequencies. The multimode configuration differs from standard single-mode periodic filters and enables flexible designs and the possibility for active control of the filter characteristics. For this proof-of-concept, we present the relevant theory and design procedures required to develop a band-stop filter that has a center frequency of $f_{\text{c}}$ = 0.8 THz and a bandwidth of $\Delta f$ = 0.07 THz. We find good agreement between theory, simulation, and experiment.}

\keywords{Terahertz, Filter, Multimode, Bandstop, Waveguide, Periodic}



\maketitle

\section{Introduction}\label{sec1}

Filters are commonly used to enhance signals in THz communication and sensing applications \cite{chen2019survey, singh2019spoof}. In a communication system, the usage of a bandstop filter improves the dynamic range. For sensing applications, the filter characteristics such as center frequency and bandwidth can be indicative of a nearby analyte. Bandstop filters are constructed using several different methods such as resonators \cite{smith2021characterization, 2005BeereTHzBandStop, cabello2022capacitively} or distributed periodic structures \cite{2017_Hollow_BG, gao2021effective, dehghanian2023demonstration}. This work presents the theory, simulation results, and experimental measurements for a new guided-wave multimode periodic bandstop filter topology that can be used at THz frequencies. The filter consists of alternating uniplanar waveguide configurations, specifically, a finite-ground plane coplanar waveguide (FGPCPW) and a coplanar stripline (CPS). For brevity, we will refer to FGPCPW as a CPW. This work primarily discusses the \textit{odd}-mode of the CPW, which is sometimes called the `parasitic' mode \cite{ghione1987coplanar}. We note that it is important to be diligent about the even/odd-mode distinction because literature commonly only discusses the `non-parasitic' \textit{even}-mode of the CPW. The multimodal nature of the CPW enables unique possibilities that are not viable with other periodic filters that use two conductors like in our previous work \cite{dehghanian2023demonstration}. For example, the presented filter can be integrated with phase shifters \cite{llamas2010mems}, hybrids \cite{llamas2009rigorous}, and diodes for reconfigurability \cite{contreras2011novel}.

Characterizing devices at THz frequencies poses unique challenges because test instrumentation is not widely available and/or is cost-prohibitive. The vector network analyzers (VNAs) that are commonly used for microwave device characterization cannot be directly used at THz frequencies. Extension modules are commercially available (albeit costly) which allow for VNAs to characterize devices up to 1.5 THz \cite{VirginiaDiodes}; however, they use rectangular waveguide feeds that are band-limited which can further increase cost when attempting to perform wideband characterization since multiple extension modules are required. Also, interfacing a rectangular waveguide with a uniplanar requires mode conversion elements which can pose a challenge due to the micrometer-scale non-planar dimensions. This work forgoes the use of VNAs and extension modules, and instead, uses thin-film photoconductive switches (PCSs) in conjunction with femtosecond optical pulses for the transmitting and receiving which provides sufficient accuracy to characterize the filter performance considered in this work \cite{grischkowsky1988capacitance}. This method is similar to THz time-domain spectroscopy (THz-TDS) \cite{lee2009principles}, and it allows for the CPS to be directly driven by a source with THz-bandwidth and resolves the signal that has traversed a device-under-test (i.e., a periodic filter in this work). Next, care must be taken when selecting the substrate thickness for uniplanar waveguides since substrate radiation can result in significant losses \cite{rutledge1983integrated}. A straightforward method to overcome this issue is to use a very thin substrate \cite{cheng1994terahertz}. In this work, the substrate is a 1 \textmu m of silicon nitride (SiN) that is mechanically suspended by a 500 \textmu m silicon (Si) frame.

Regarding filter specifications, this work is a proof-of-concept and we select a center frequency of $f_{\text{c}}$ = 0.8 THz and a bandwidth of $\Delta f$ = 0.07 THz. We also select to use 20 unit cells which resulted in a stopband rejection of 15-20 dB. These specifications were selected because we were confident that our experiments could clearly resolve device characteristics in this frequency window. The presented equations and theory can be scaled to higher frequencies.

Section \ref{sec:theory} discusses the theory for the periodic filter in terms of modes and characteristic impedances using ABCD matrices. Section \ref{sec:methods} presents the methods used for fabrication and performing the experiment. Section \ref{sec:results} presents the experimental results and discusses the measured spectrum. Section \ref{sec:conclusion} is the conclusion. Appendix \ref{sec:AppendixA_Reconfig} illustrates an example that demonstrates that the multimode property can be exploited to convert the presented band-stop filter into a band-pass filter.

\section{Design and Theory}
\label{sec:theory}

\subsection{Design}
\begin{figure}[H]
    \centering
    \includegraphics[width=\linewidth]{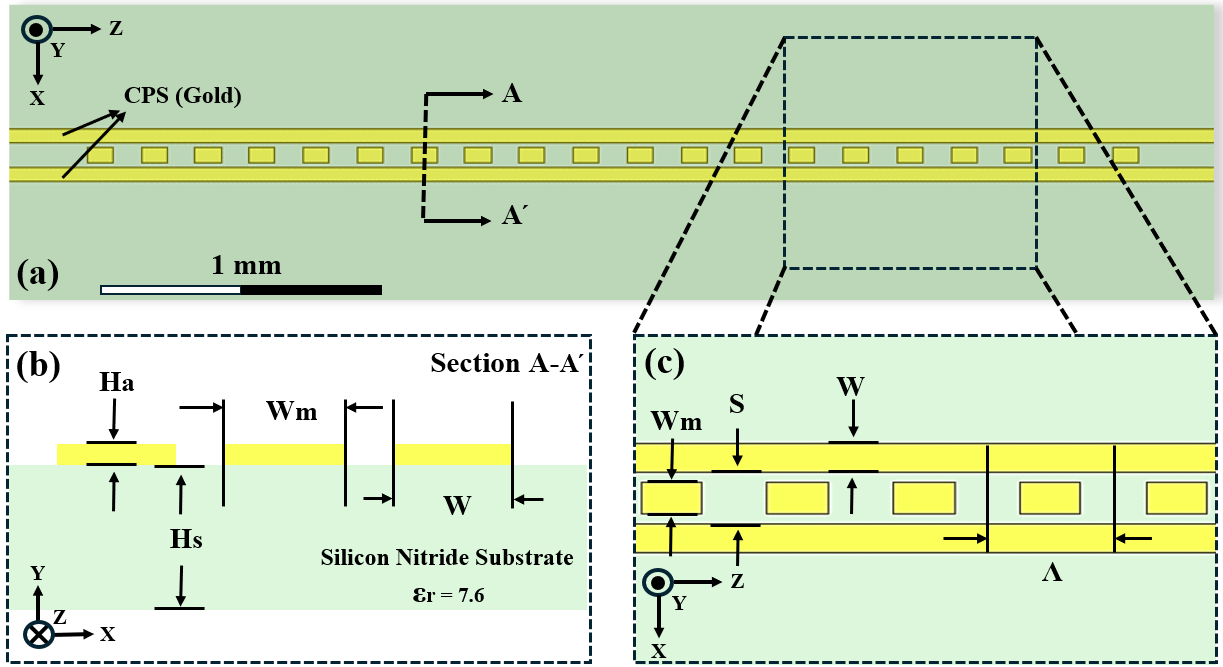}
    \caption{(a) Overall structure. (b) Cross section of the CPW sections. (c) Definition of the parameters associated with the unit cell.}
    \label{fig:structure}
\end{figure}

The filter is constructed from alternating transmission line configurations with different characteristic impedances. Specifically, we alternate between a CPS and a CPW (see Fig. \ref{fig:structure}a). Figure \ref{fig:structure}b illustrates the transverse cross-section of the CPW section. Figure \ref{fig:structure}c illustrates a section of the filter that clarifies the unit cell. The conductor width is $W$ = 45 \textmu m and the separation is $S$ = 80 \textmu m. These parameters were selected because they exhibit reasonable loss characteristics at THz frequencies \cite{smith2019demonstration}. The width of the central conductor, $W_m$, is used to vary the characteristic impedance of the CPW mode which is investigated in Section \ref{sec:modes}. For the experiment, we selected $W_m$ = 45 \textmu m which corresponds to a stopband attenuation of approximately -20 dB with $N = 20$ filter periods. As mentioned, the filter is fabricated on a thin ($H_s$ = 1 \textmu m) SiN substrate to minimize radiation loss and dispersion at THz frequencies \cite{cheng1994terahertz}. The conductor thickness was selected to be $H_a$ = 200 nm to have reasonable conductor loss while remaining compatible with our transmitter and receivers' bonding method \cite{Rios2015_bowtie_PCA}. If other bonding methods are used, then thicker conductors can be used for lower conductor loss. The grating period is selected to be $\Lambda$ = 173 \textmu m which corresponds to a filter center frequency of $f_c$ = 0.8 THz (discussed later). 

\subsection{Simulation details}
Full-wave simulations (ANSYS HFSS 2024 R2) are used for obtaining mode impedances, S-parameters, and for producing field illustrations. The material constitutive parameters are as follows: the conductors are modeled as gold (Au) with $\sigma_{\text{Au}}$(0.8 THz) = 2.16 $\times$ 10$^7$ S/m \cite{lucyszyn2004investigation}, the substrate is silicon nitride (SiN) and modeled with $\varepsilon_{r,\text{SiN}} = 7.6$, $\mu_{r,\text{SiN}} = 1$, and $\tan \delta_{\varepsilon, \text{SiN}} = 0.00526$ \cite{Cataldo_Silicon_nitride_properties_2012}.

\subsection{Modes and coupling}
\label{sec:modes}
\begin{figure}[H]
    \centering
    \includegraphics[width=0.9\linewidth]{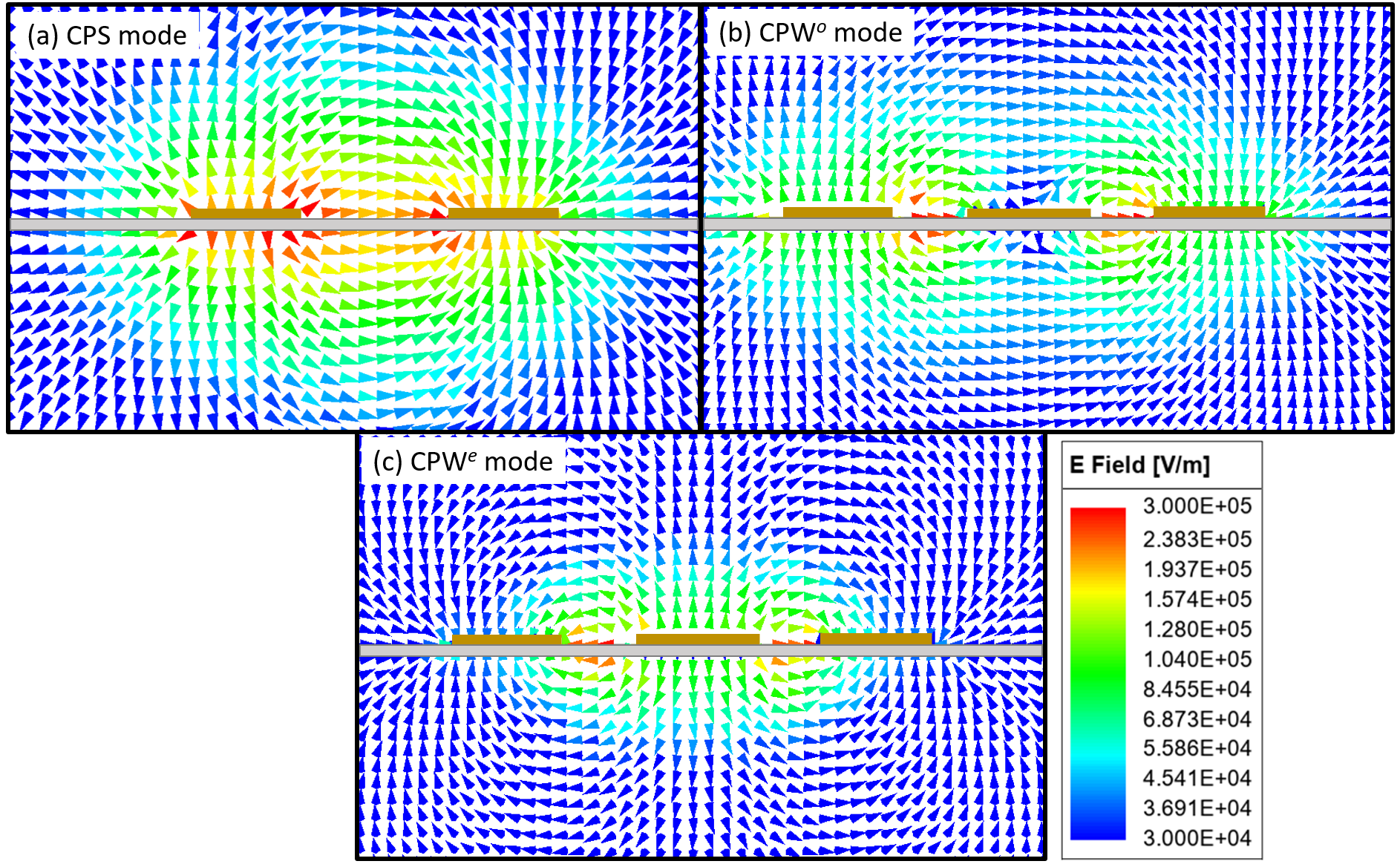}
    \caption{Mode illustrations. (a) CPS mode. (b) CPW odd mode. (c) CPW even mode. The conductors and substrates are overlaid for visibility, their thicknesses are not to scale ($\approx$10$\times$ thinner than visualized).}
    \label{fig:mode_illustrations}
\end{figure}

\begin{figure}[H]
    \centering
    \includegraphics[width=0.9\linewidth]{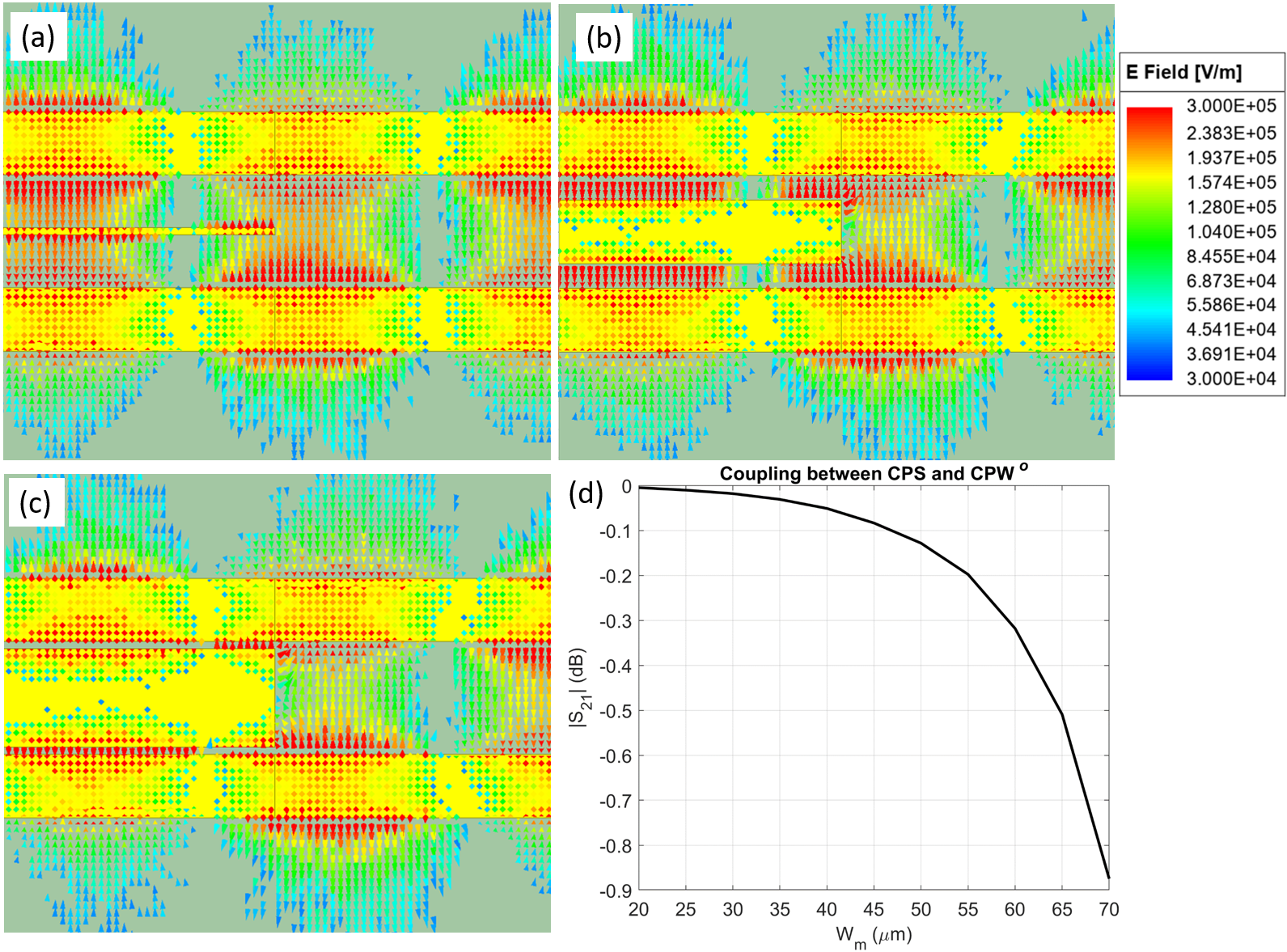}
    \caption{CPS-to-CPW$^o$ mode coupling at 0.8 THz. (a) $W_m$ = 5 \textmu m. (b) $W_m$ = 45 \textmu m. (c)  $W_m$ = 70 \textmu m. (d) $|S_{21}^{o}|^2$ for CPS-to-CPW$^o$ vs $W_m$.}
    \label{fig:mode_coupling}
\end{figure}

The CPW consists of three conductors and therefore it supports an even and an odd mode. We use superscripts to differentiate the odd and even modes for the CPW, e.g., CPW$^o$ and CPW$^e$. The two modes have different characteristic impedances: $Z_{\text{CPW}}^{o}$ and $Z_{\text{CPW}}^{e}$. As previously mentioned, the $\text{CPW}^{e}$ mode is typically the desired mode, whereas the $\text{CPW}^{o}$ mode is generally considered parasitic and is suppressed by short-circuiting the outer conductors. However, in this work, the $\text{CPW}^{e}$ mode is of secondary importance and we focus on the  $\text{CPW}^{o}$ mode. As an aside, we note that it is possible to excite the $\text{CPW}^{e}$ mode to create interesting devices. An example is given in Appendix \ref{sec:AppendixA_Reconfig} where the bandstop filter is converted to a bandpass filter by exploiting this property. Continuing on, the CPS has two conductors thus it only supports a single mode and characteristic impedance, $Z_{\text{CPS}}$. Figure \ref{fig:mode_illustrations} illustrates these three modes in the transverse plane. Investigation of Fig. \ref{fig:mode_illustrations}(a,b) reveals that the CPS mode resembles CPW$^o$ mode and significant modal coupling is expected. Figure \ref{fig:mode_coupling} qualitatively and quantitatively illustrates the impact of a CPW$^o$-CPS interface. Figure \ref{fig:mode_coupling}(a-c) illustrates the electric field vector at 0.8 THz for several widths of the central conductor, $W_m$, where minimal disruption is observed at the CPW$^o$-CPS interface. To quantify the transmission at the interface we plot $|S_{21}^{o}|^2$ (transmission between CPW$^o$ and CPS) vs $W_m$ in Fig. \ref{fig:mode_coupling}(d). We do not plot $|S_{21}^{e}|^2$ (transmission between CPW$^e$ and CPS) vs $W_m$, since it is less than -60 dB which is expected when the CPW is symmetric \cite{ribo1999circuit}. From Fig. \ref{fig:mode_coupling}(d) it is apparent that there is good coupling between CPS and CPW$^o$ modes and thus the configuration can form a building block for a periodic filter. We note that the central conductor does not need to be rectangular or symmetric, but the presented design and theory would become more complex -- these are not evaluated in this paper.  Lastly, the transition between the CPS and CPW could include a shunt admittance to model parasitic effects, but we found that it had negligible impact on the periodic filter investigated in this work.

\subsection{Filter characteristics}

The center frequency of the periodic filter is dependent on the period ($\Lambda$) by the following equation:
\begin{equation}
f_c = \frac{c}{2 \Lambda \sqrt{\varepsilon_{re}}},
\label{eqn:fc}
\end{equation}

\noindent where $c$ is the speed of light and $\varepsilon_{re}$ is the mean effective relative permittivity of the modes which was obtained from full-wave simulation and is given by $\varepsilon_{re} \approx 1.2$, which results in $f_c$ = 0.8 THz when $\Lambda$ = 173 \textmu m. Next, the fractional bandwidth is controlled by the characteristic impedance's ($Z_{\text{CPW}}^{o}$ and $Z_{\text{CPS}}$) as follows:

\begin{equation}
    \frac{\Delta f}{f_c} = \frac{4}{\pi} \sin^{-1} \left( \frac{Z_{\text{CPS}}-Z_{\text{CPW}}^{o}}{Z_{\text{CPS}}+Z_{\text{CPW}}^{o}} \right).
    \label{eqn:frac_bandwidth}
\end{equation}

\begin{figure}[H]
    \centering
    \includegraphics[width=\linewidth]{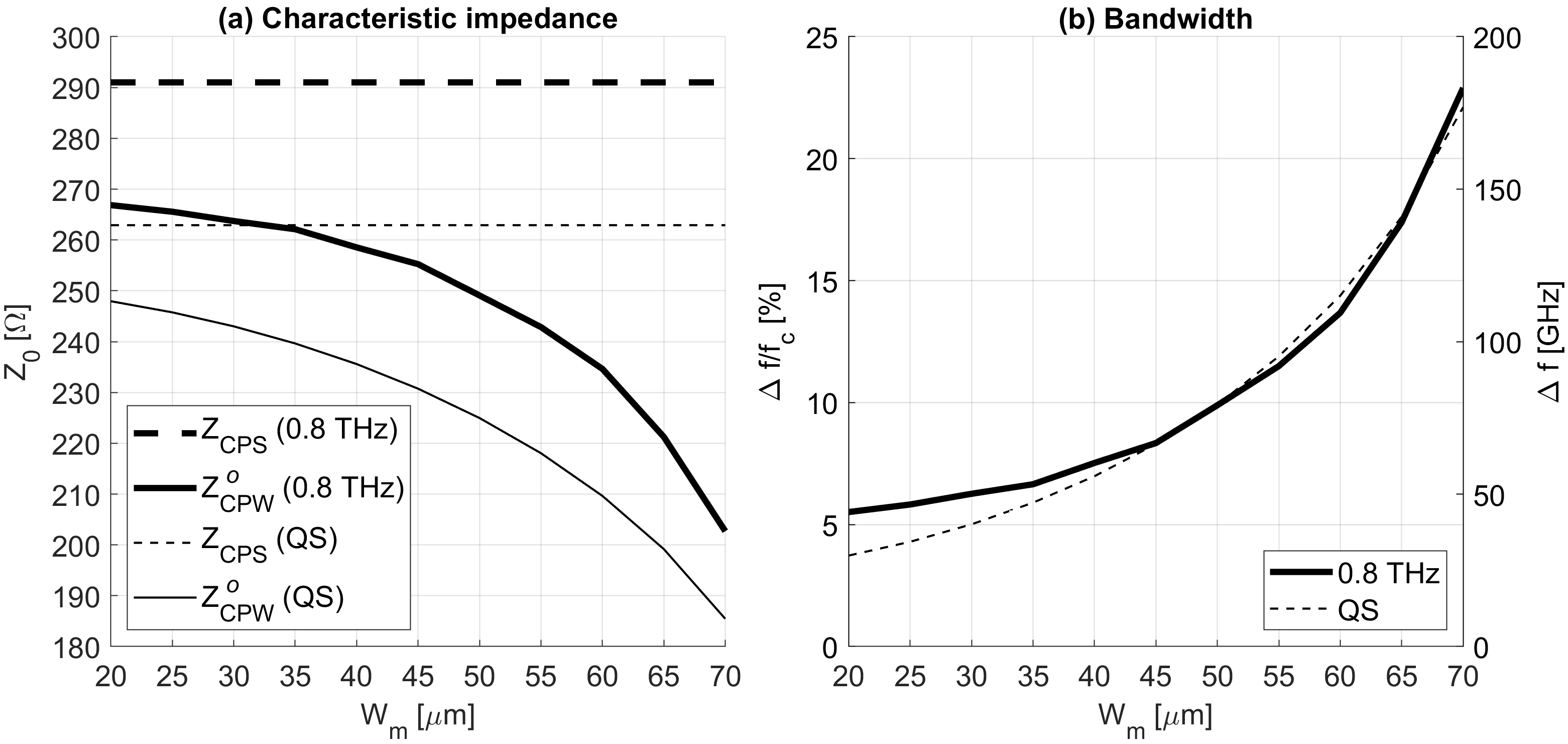}
    \caption{(a) Characteristic impedance using QS expressions and full-wave simulation at 0.8 THz with $S$ = 80 \textmu m, $W$ = 45 \textmu m, $H_s$ = 1 \textmu m, $H_a$ = 200 nm, and $\varepsilon_r = 7.6$. The impedance at 0.8 THz with $W_m$ = 45 \textmu m are $Z_{\text{CPS}} = 291 \Omega$ and $Z_{\text{CPW}}^o = 254 \Omega$. (b) The corresponding fractional bandwidth from (\ref{eqn:frac_bandwidth}) and bandwidth where $f_c = 0.8$ THz. At $W_m$ = 45 \textmu m, the bandwidth is 0.07 THz.}
    \label{fig:mode_Z0}
\end{figure}

During the initial design it is helpful to estimate the device parameters and characteristic impedances, this approximation can be performed assuming from the quasi-static (QS) expressions. For the CPW sections, the QS characteristic impedance of the odd-mode is given by \cite{ghione1987coplanar}:  

\begin{equation}
    Z_{\text{CPW}}^{o} = \frac{120 \pi}{\sqrt{\varepsilon_{re}}}\frac{K(k_{CPW})}{K(k_{CPW}')}, \quad k_{CPW} = \frac{S}{S+2W} \sqrt{\frac{1-\left(\frac{W_m}{S}\right)^2}{1-\left(\frac{W_m}{S+2W}\right)^2}},
    \label{eqn:ZCPW}
\end{equation}

\noindent where $k_{CPW}' = \sqrt{1-k_{CPW}^2}$ and $K(\cdot)$ is the complete elliptic integral of the first kind. The QS characteristic impedance of the CPS sections is given by \cite{ghione1984analytical}:  

\begin{equation}
    Z_{\text{CPS}} = \frac{120 \pi}{\sqrt{\varepsilon_{re}}}\frac{K(k_{CPS})}{K(k_{CPS}')}, \quad k_{CPS} = \frac{S}{S+2W},
    \label{eqn:ZCPS}
\end{equation}

\noindent where $k_{CPS}' = \sqrt{1-k_{CPS}^2}$. Practical insight is obtained from (\ref{eqn:ZCPW}) and (\ref{eqn:ZCPS}), specifically as $W_m \rightarrow 0$, then $k_{CPW} = k_{CPS}$ which leads to $Z_{\text{CPW}}^{o} = Z_{\text{CPS}}$. From (\ref{eqn:frac_bandwidth}) this implies that the fractional bandwidth reduces to zero, which is aligned with the expectation of connecting two identical sections of transmission line (i.e., the CPW with $W_m$ = 0 and a CPS). Figure \ref{fig:mode_Z0}a plots the characteristic impedance of the modes using the QS expressions and at 0.8 THz, which was obtained from full-wave simulation. The QS and 0.8 THz traces exhibit similar characteristics but are shifted in value, which is expected as the characteristic impedance of CPW and CPS is frequency dependent \cite{shih1982analysis, phatak1990dispersion}. As noted, the QS expressions of (\ref{eqn:ZCPW}) and (\ref{eqn:ZCPS}) can be used as a starting point for the initial design, but the final design should use full-wave simulation to obtain the characteristic impedance at the design frequency. The bandwidth is obtained using (\ref{eqn:frac_bandwidth}) and is plotted in Fig. \ref{fig:mode_Z0}b, where a bandwidth of 0.07 THz is predicted for the experimental structure with $W_m$ = 45 \textmu m.

To summarize, the design procedure begins by selecting a desired center frequency using (\ref{eqn:fc}). Next, the effective relative permittivity, $\varepsilon_{re}$, is obtained using full-wave simulation, but when using thin substrates (as in this work) then $\varepsilon_{re} \approx 1$. Next, the conductor width, $W$, should be selected to have desirable attenuation characteristics at THz frequencies such as $W = $ 30 \textmu m. The conductor separation, $S$, is then selected to obtain a desired characteristic impedance for the feedline, $Z_{CPS}$ using (\ref{eqn:ZCPS}) (or full-wave simulation). The central conductor width, $W_m$, is selected to obtain $Z_{CPW}^o$ using (\ref{eqn:ZCPW}) (or full-wave simulation) which produces the specified fractional bandwidth given by (\ref{eqn:frac_bandwidth}).

\subsection{Periodic Response}
\subsubsection{Theory}
The periodic filter is constructed by cascading, $N$, unit cells that consist of two quarter wavelength ($\lambda/4$) sections of transmission lines with different characteristic impedances. The unit cells are modeled using ABCD matrices. To simplify the analysis, it is best to construct the unit cell's ABCD matrix using a CPS of length $\Lambda/4$, a CPW$^o$ of length $\Lambda/2$ , and a CPS of length $\Lambda/4$. This enables a straightforward method to calculate the total ABCD matrix by raising the unit cell's ABCD matrix to the power of $N$. The ABCD matrix for a unit cell (see Fig. \ref{fig:structure}) is given by the following:
\begin{equation}
\begin{split}
\begin{bmatrix}
A_{cell} & B_{cell} \\
C_{cell} & D_{cell} 
\end{bmatrix}  =
& \begin{bmatrix}
\text{cos}( \beta_{\text{CPS}} \Lambda/4) & j Z_{\text{CPS}} \text{sin} (\beta_{\text{CPS}} \Lambda/4) \\
j (1/Z_{\text{CPS}}) \text{sin}( \beta_{\text{CPS}} \Lambda/4 )& \text{cos} (\beta_{\text{CPS}} \Lambda/4 )
\end{bmatrix} \\
\cdot & \begin{bmatrix}
\text{cos}( \beta_{\text{CPW}}^o \Lambda/2 )& j Z_{\text{CPW}}^{o} \text{sin} (\beta_{\text{CPW}}^o \Lambda/2) \\
j (1/Z_{\text{CPW}}^{o}) \text{sin}( \beta_{\text{CPW}}^o \Lambda/2 )& \text{cos} (\beta_{\text{CPW}}^o \Lambda/2 )
\end{bmatrix} \\
\cdot & \begin{bmatrix}
\text{cos}( \beta_{\text{CPS}} \Lambda/4) & j Z_{\text{CPS}} \text{sin} (\beta_{\text{CPS}} \Lambda/4) \\
j (1/Z_{\text{CPS}}) \text{sin}( \beta_{\text{CPS}} \Lambda/4 )& \text{cos} (\beta_{\text{CPS}} \Lambda/4 )
\end{bmatrix}.
\end{split}
\label{eqn:ABCD_cell}
\end{equation}
The ABCD matrix for $N$ sections is given by:
\begin{equation}
\begin{bmatrix}
A & B \\
C & D 
\end{bmatrix} =
\begin{bmatrix}
A_{cell} & B_{cell} \\
C_{cell} & D_{cell} 
\end{bmatrix}^N
\label{eqn:ABCD_total}
\end{equation}

\begin{figure}[H]
    \centering
    \includegraphics[width=0.5\linewidth]{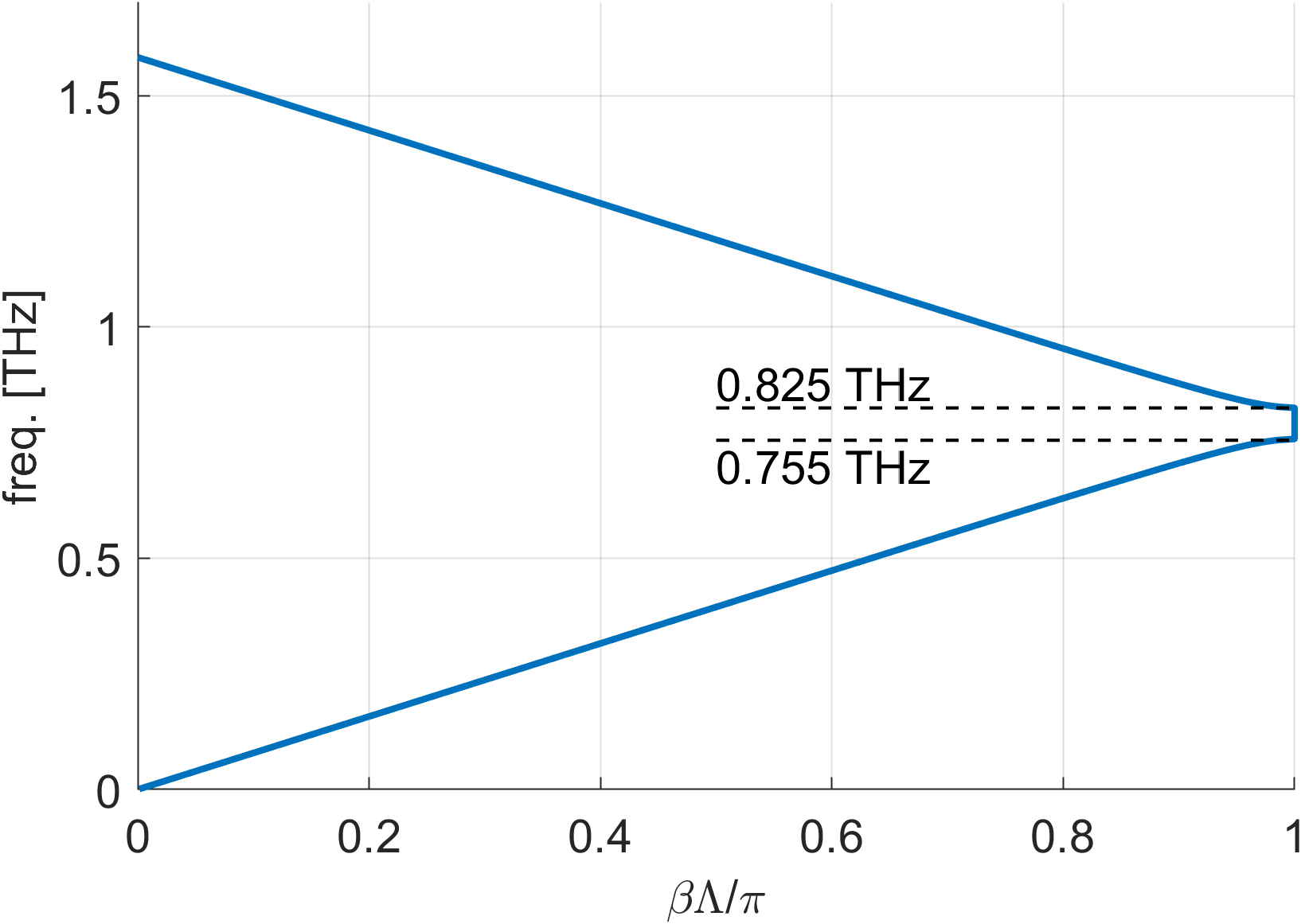}
    \caption{Dispersion diagram for the unit cell.}
    \label{fig:dispersion}
\end{figure}

The dispersion characteristics of the filter are obtained from (\ref{eqn:ABCD_cell}) using the same methods presented in \cite{dehghanian2023demonstration}. The result of this process is plotted in Fig. \ref{fig:dispersion} where a 0.07 THz stopband is observed.

Lastly, the S-parameters are calculated from the ABCD matrix using standard methods \cite{pozar2021microwave}:
\begin{equation}
    S_{11} = \frac{A+B/Z_{\text{CPS}}-CZ_{\text{CPS}}-D}{A+B/Z_{\text{CPS}}+CZ_{\text{CPS}}+D} \quad \text{and} \quad S_{21} = \frac{2}{A+B/Z_{\text{CPS}}+CZ_{\text{CPS}}+D}
\label{eqn:ABCD_to_S}
\end{equation}

\subsubsection{Simulation}
\begin{figure}[H]
    \centering
    \includegraphics[width=1\linewidth]{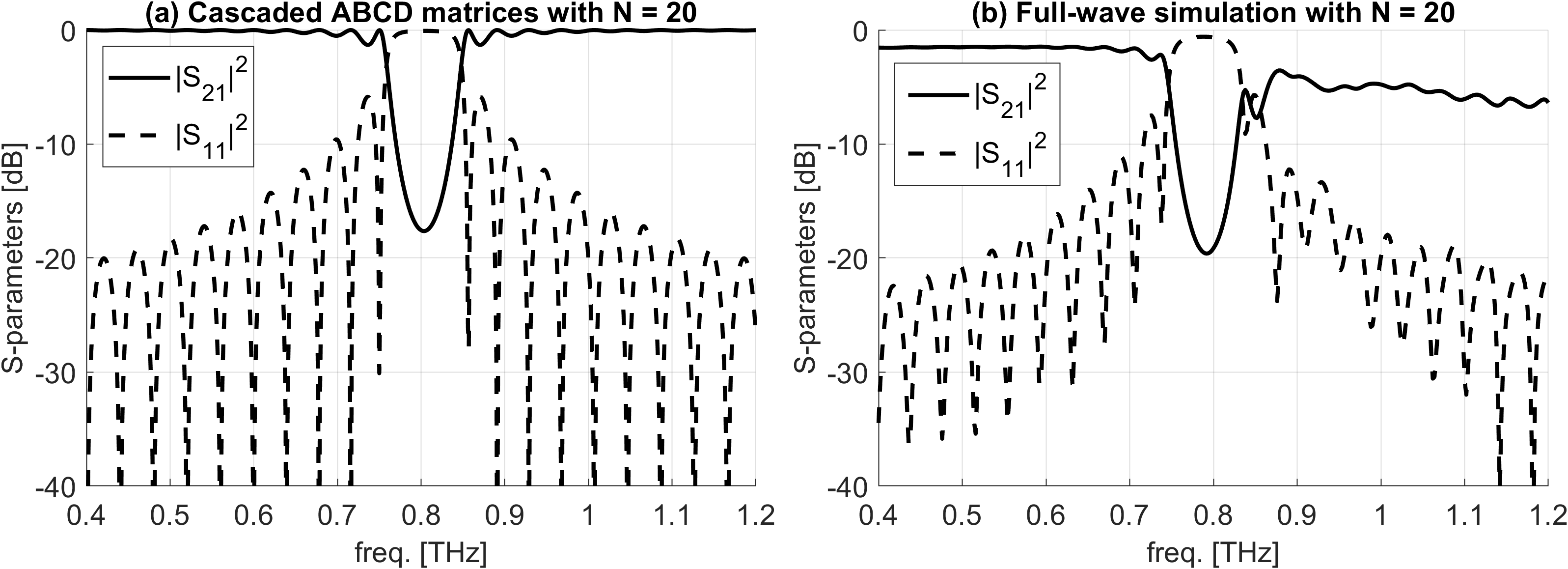}
    \caption{(a) S-parameters obtained from cascading twenty unit cells ($N = 20$) using ABCD matrices. The unit cell is included as an inset. (b) S-parameters obtained from cascading twenty unit cells (N $ = 20$) using ANSYS HFSS.}
    \label{fig:S-params_vs_freq}
\end{figure}

A full-wave simulation was also used to characterize the total response of the filter with $N=20$ near the design frequency. These simulations account for the full-wave effects that are not modeled by the ABCD matrices. Figure \ref{fig:S-params_vs_freq}a plots the S-parameters for $N=20$ unit cells that were obtained using (\ref{eqn:ABCD_cell}), (\ref{eqn:ABCD_total}), and (\ref{eqn:ABCD_to_S}) where the characteristic impedances were obtained from Fig. \ref{fig:mode_Z0} at 0.8 THz. Figure \ref{fig:S-params_vs_freq}b plots the S-parameters for the periodic filter obtained from HFSS. Comparing Fig. \ref{fig:S-params_vs_freq}a and Fig. \ref{fig:S-params_vs_freq}b shows that the ABCD model provides a good first approximation. The main difference observed between theory and simulation is the introduction of passband insertion loss and a difference in the stopband rejection. At frequencies below the stopband, there is approximately 1 dB of insertion loss which originates from conductor loss. This effect could be included in the ABCD matrix model if lossy transmission line models were used. Above the stopband, the insertion loss increases to 4--6 dB which originates from diffractive grating radiation \cite{cheben_subwavelength_2018} which is not straightforward to model using ABCD matrices; thus, full-wave simulations should be used to obtain a comprehensive understanding of device performance.

\begin{figure}[H]
    \centering
    \includegraphics[width=1\linewidth]{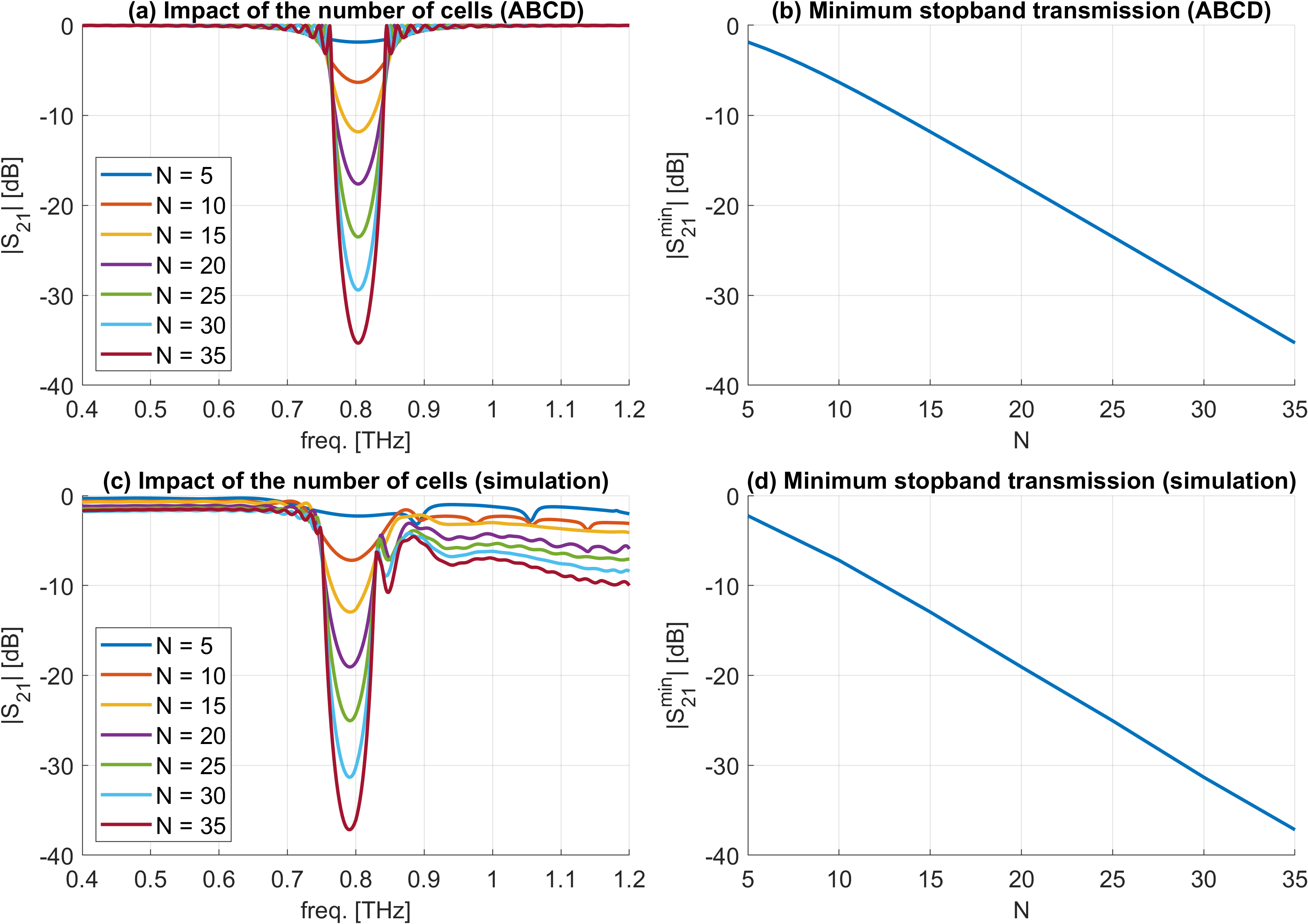}
    \caption{(a) Transmission response of the filter using ABCD matrices for several different numbers of unit cells, N. (b) Minimum stopband transmission using ABCD matrices plotted against the number of unit cells, N. (c) Transmission response of the filter using ANSYS HFSS for several different numbers of unit cells, N. (d) Minimum stopband transmission using ANSYS HFSS plotted against the number of unit cells, N.}
    \label{fig:S-params_vs_N}
\end{figure}

Figure \ref{fig:S-params_vs_N}(a,b) use (\ref{eqn:ABCD_total}) and (\ref{eqn:ABCD_to_S}) to plot the impact of the number of unit cells on the stopband rejection. From Fig. \ref{fig:S-params_vs_N}b, it is found that while $N > 10$, the minimum stopband transmission is given by $|S_{21}^{\text{min}}|$ [dB] = -1.18$N$ + 5.90. Figure \ref{fig:S-params_vs_N}(c-d) plots the same information, but uses full-wave simulation instead of ABCD matrices. The minimum stopband transmission is given by $|S_{21}^{\text{min}}|$ [dB] = -1.21$N$ + 5.00 which is similar to the ABCD method; however, diffractive grating radiation is observed and reduces the transmission above the stopband by approximately -0.2$N$ in dB at 1.1 THz.

\section{Methods}
\label{sec:methods}

\subsection{Fabrication}

\begin{figure*}[t]
  \centering
  \includegraphics[width=\textwidth]{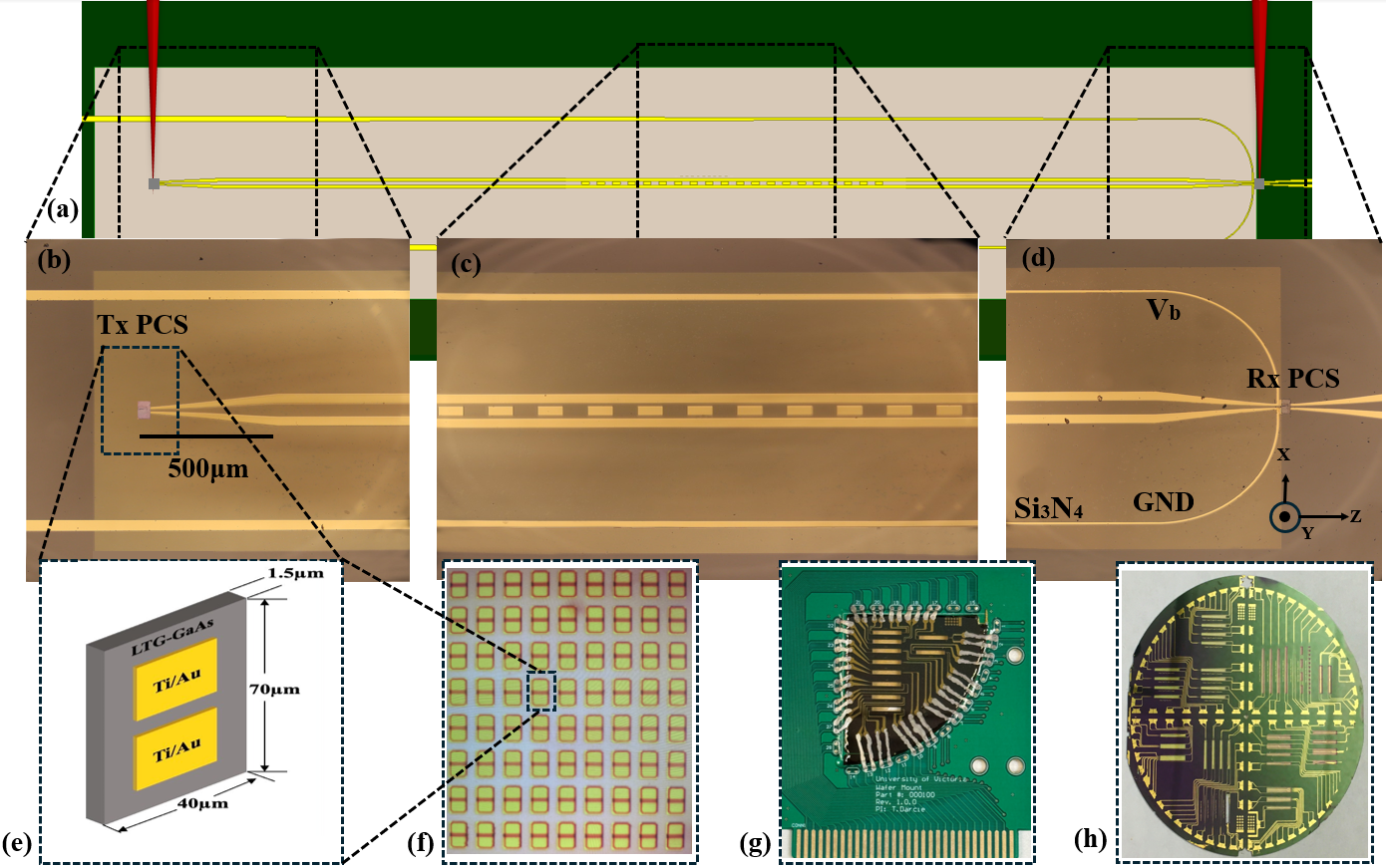}
  \caption{(a) The complete integrated system fabricated on a thin SiN membrane. (b) The transmitter section, consisting of an LT-GaAs PCS, 200 nm thick gold CPS TL on a 1 \textmu m SiN layer. (c) A microscopic view of the periodic filter. (d) The receiver section, including the DC block and bias lines for the PCS and lock-in amplifier connection. (e) Thin films of LT-GaAs with dimensions of 70 \textmu m $\times$ 40 \textmu m $\times$ 1.8 \textmu m, used in the PCS arrays. (f) The PCS arrays after fabrication, used in both transmitter and receiver sections. (g) A quarter wafer mounted on a PCB for alignment and connection to the measurement setup, ensuring precise signal transmission. (h) Fabricated circuits on the wafer before being diced into four quarters. }
  \label{fig:Fab}
\end{figure*}

The fabricated filter (Fig. \ref{fig:Fab}) consists of a patterned metallization on a thin SiN substrate that uses low-temperature grown gallium arsenide (LT-GaAs) photoconductive switches (PCSs) for signal generation and detection. The substrate is fabricated by depositing a 1 \textmu m layer of SiN on a 500 \textmu m silicon wafer. The metallization was subsequently deposited using a combination of ultraviolet (UV) lithography and electron-beam physical vapor deposition (EBPVD). The metallization used a 10 nm titanium (Ti) adhesion layer and a 200 nm gold (Au) layer. Lastly, the Si substrate is masked using UV lithography and selectively etched using potassium hydroxide (KOH) resulting in a suspended SiN window with the desired conductor pattern.

The LT-GaAs PCSs (Fig. \ref{fig:Fab}e) were constructed using a method similar to the technique described in \cite{Rios2015_bowtie_PCA}. In short, an epitaxial lift-off method is used \cite{yablonovitch1990van}. First, using molecular beam epitaxy (MBE), a 2 \textmu m layer of LT-GaAs is deposited on a 50 nm aluminum arsenide (AlAs) layer on a 500 nm semi-insulating GaAs substrate. The AlAs layer is a sacrificial layer that is selectively etched using a 10\% hydrofluoric (HF) acid etchant. Before etching the AlAs layer, a grid (Fig. \ref{fig:Fab}f) of individual PCS devices is prepared by patterning the LT-GaAs surface with Ti/Au (10 nm/200 nm) via RF sputtering. Next, the LT-GaAs surface contact pairs are masked with a photoresist using UV lithography, then the exposed regions of the LT-GaAs surface are etched using citric acid and hydrogen peroxide to produce discrete PCSs. Next, the surface is coated in an etch resistant wax (Apiezon W) then submerged in a HF acid bath until the AlAs is fully dissolved and the LT-GaAs layer is released. Next, a thin layer of LT-GaAs is dissolved from the backside using the citric acid and hydrogen peroxide etchant. Lastly, the individual PCS devices are obtained after dissolving the Apiezon W wax using trichloroethylene (TCE). The resultant discrete PCSs can then be placed on uniplanar circuits and connected using Van der Waals bonding \cite{yablonovitch1990van}.

\subsection{Experiment}
\begin{figure}[H]
    \centering
    \includegraphics[width=0.7\linewidth]{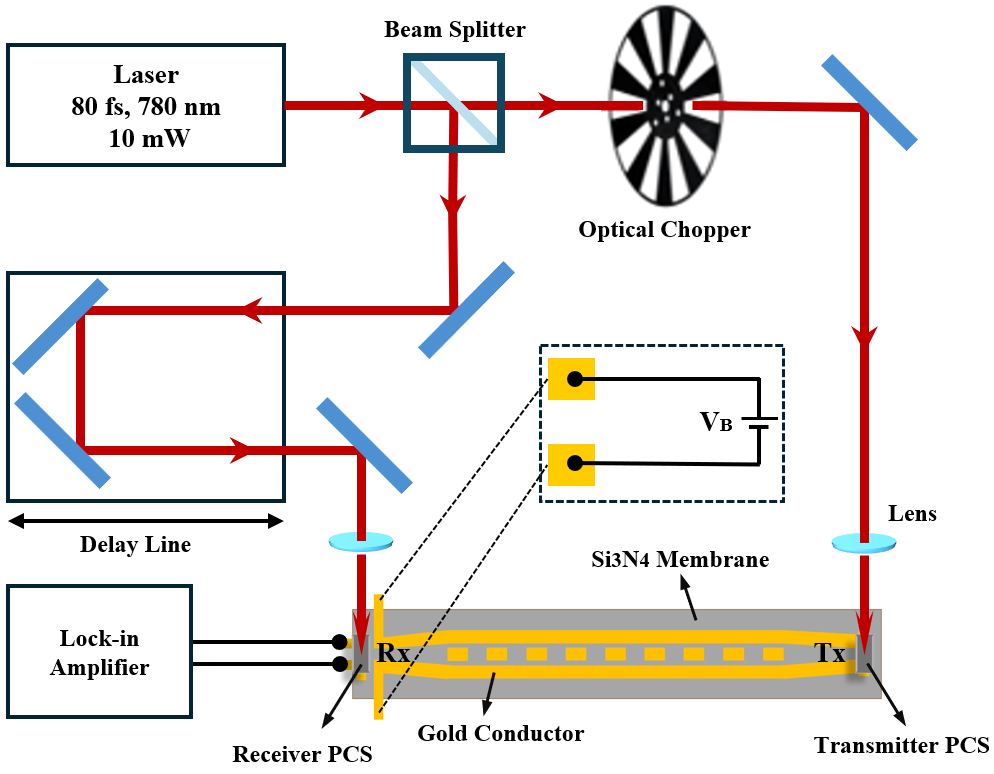}
    \caption{Optical setup used for characterizing the filter response. The blue rectangles are mirrors.}
    \label{fig:experiment}
\end{figure}

We characterize the response of the periodic filter at THz frequencies using a modified THz time-domain spectrometer (Fig. \ref{fig:experiment}). The THz-bandwidth signal originates from a DC biased (V$_{\text{B}}$ = 24 V) PCS that is illuminated by a femtosecond laser (CALMAR Carmel) with a pulse width of $\tau_{\text{p}}$ = 80 fs, repetition rate of f$_{\text{rep}}$ = 80 MHz, and an average optical power of P$_{\text{opt}}$ = 10 mW. The biased PCS and transient photoconductivity result in a subpicosecond current pulse that drives the feedlines and the periodic filter. The transient signal propagates through the periodic filter and is recovered by sampling the current via lock-in detection (Stanford Instruments SR830) generated by the combination of the incident voltage and receiver PCS at different times that are controlled by a mechanical delay line. The lock-in amplifier is referenced to the optically chopped (f$_{\text{chop}}$ = 1 KHz) transmitter beam. The result of the experiment is a transient signal that contains frequency components up to THz frequencies that are obtained by applying the discrete Fourier transform (DFT). It is important to state that our experimental methodology is not akin to a standard vector network analyzer (VNA) measurement that directly exports the devices' S-parameters when properly calibrated. Our signal source is a finite duration transient pulse; thus, it exhibits an inherent roll-off with frequency. When interpreting the experimental results, the transmission is obtained by comparing the responses of a reference ($W_m$ = 0 \textmu m) and the filter ($W_m$ = 45 \textmu m).

\section{Results and Discussion}
\label{sec:results}

\begin{figure}[H]
    \centering
    \includegraphics[width=\linewidth]{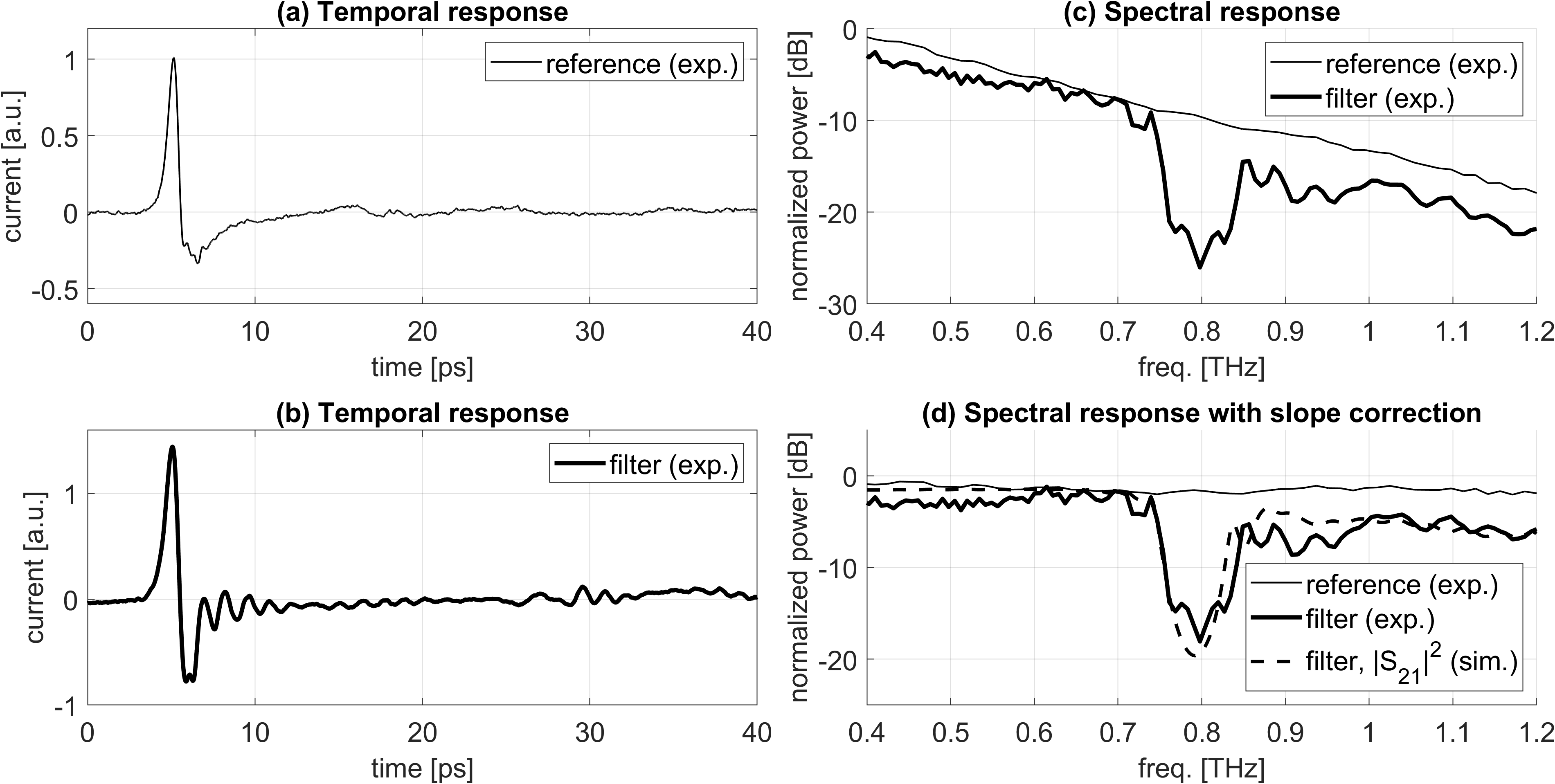}
    \caption{Experimental results. (a) Temporal response for a reference ($W_m$ = 0 \textmu m). (b) Temporal response a for periodic filter ($W_m$ = 45 \textmu m) spectrally normalized to the reference. (c) Spectral response for a reference and periodic filter obtained via DFT. (d) Spectral response with 20 dB/THz slope correction alongside a full-wave $|S_{21}|^2$ simulation results from Fig. \ref{fig:S-params_vs_freq}b.}
    \label{fig:exp_results}
\end{figure}

Figure \ref{fig:exp_results} plots the experimental results for a reference structure and the periodic filter illustrated in Fig. \ref{fig:Fab}(b-d). The temporal response is obtained using the experiment shown in Fig. \ref{fig:experiment} and the spectral response is obtained by applying the DFT to the temporal response. The temporal and spectral responses are normalized. The spectral response in Fig. \ref{fig:exp_results}c illustrates a stopband centered at 0.8 THz with a bandwidth of $\approx$0.1 THz which aligns with the theory (Fig. \ref{fig:S-params_vs_freq}a) and full-wave simulations (Fig. \ref{fig:S-params_vs_freq}b). We reiterate that there is an expected frequency decay due to the transient excitation and detection methods. In Fig. \ref{fig:exp_results}c this appears as an approximately linear slope (on a semi-log graph) of -20 dB/THz. The slope is not always linear and is dependent on many parameters but primarily the optical pulse duration and the PCS carrier lifetime \cite{duvillaret2001analytical, garufo2018norton, zhang2024time}. To aid in interpreting the results, we add the 20 dB/THz slope to the experimental data and plot it alongside the full-wave simulation results in Fig. \ref{fig:exp_results}d. This slope is a characteristic of the transmitter and receiver; it does not mean that the filter exhibits a 20 dB/THz insertion loss. Applying the inverse slope correction to the spectral response flattens the spectrum to allow for better visual comparison against full-wave simulation ($|S_{21}|^2$). We find good relative agreement between the simulation and experimental results where the center frequency and bandwidth are in close agreement. We also observe the increased insertion loss above the stopband that is associated with diffractive grating radiation. The experimental bandwidth is also marginally wider than predicted, which likely originates from manufacturing differences.

\section{Conclusion}
\label{sec:conclusion}

We presented a multimode periodic band-stop filter based on alternating CPS and CPW sections. The filter had a center frequency of $f_{\text{c}}$ = 0.8 THz and a bandwidth of $\Delta f \approx$ 0.1 THz. The theory was based on cascading ABCD matrices that represented the CPS and the odd-mode of the CPW. The theory, simulation, and experiment are in agreement with each other. Lastly, to demonstrate a benefit of the multimode periodic filter, we provided a simulated example that used the even-mode to convert the band-stop filter into a band-pass filter by introducing a short circuit between the conductors.

\backmatter

\bmhead{Acknowledgments}

This work was supported by an NSERC Discovery Grant. The authors thank 4D LABS at Simon Fraser University for the fabrication of the CPS waveguides and the thin membrane, and also the Centre for Advanced Materials and Related Technology (CAMTEC) at the University of Victoria for providing Nanofab facilities for the fabrication of the PCS devices.

\bmhead{Author Contributions} A. D. designed and simulated the device. M.H. assisted with device fabrication, assembly, and experiments. T.D. provided insight into device performance. L.S. developed the theory and wrote the manuscript text.

\bmhead{Funding} Natural Sciences and Engineering Research Council of Canada (RGPIN-2022-03277).

\bmhead{Data availability} There are no supplementary materials, and the data is available upon reasonable request.

\section*{Declarations}

\bmhead{Ethics Approval} Not applicable.

\bmhead{Competing Interests} The authors declare that they have no competing interests.

\begin{appendices}

\section{Multimode Band-pass Filter Example}\label{sec:AppendixA_Reconfig}

\begin{figure}[H]
    \centering
    \includegraphics[width=\linewidth]{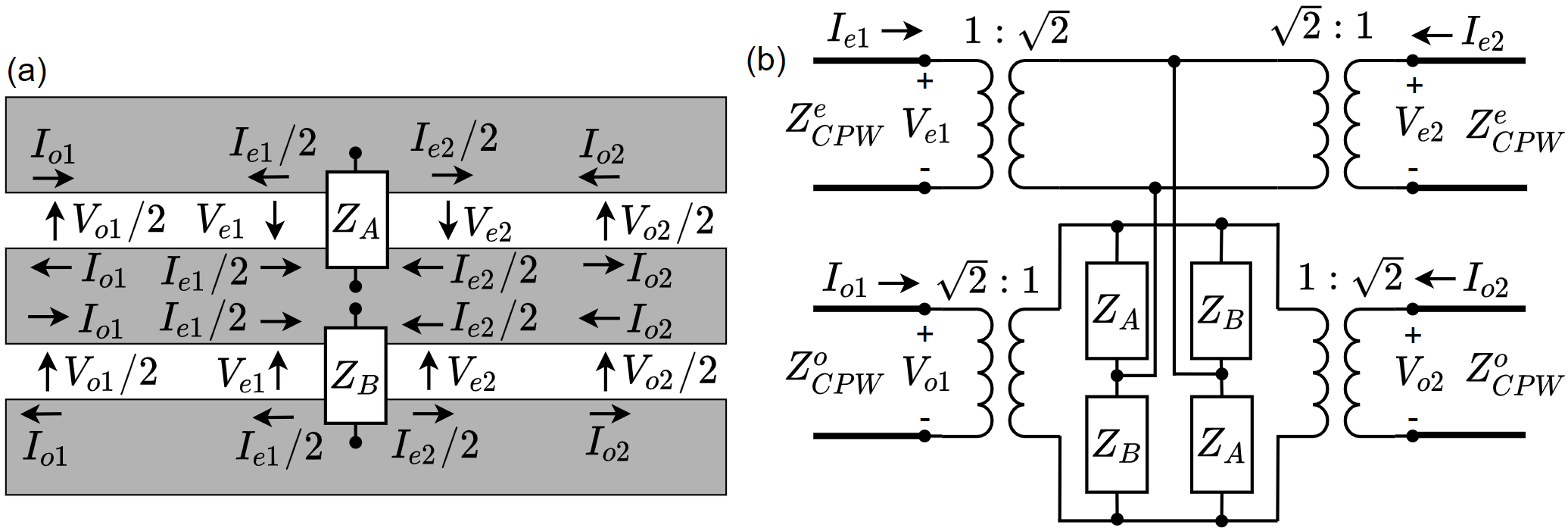}
    \caption{Illustration of the even and odd modes for a CPW with an impedance placed between the conductors.}
    \label{fig:CPW_modes_circuit}
\end{figure}

The filter presented in this work offers unique opportunities that are achieved by modifying the excitation of the even mode for the CPW sections. The bulk of this paper ignores the even mode because it is not excited; however, excitation can be achieved by considering the equivalent circuit of Fig. \ref{fig:CPW_modes_circuit} \cite{ribo1999circuit, contreras2011novel}. So far $Z_A = Z_B = \infty$, which implies that the even mode cannot be excited. Alternatively, if a short circuit ($Z_A = 0$) was placed between the conductors and the other left open ($Z_B = \infty$), then the even mode would be excited. This additional short-circuit can be used to convert the band-stop filter to a band-pass filter. 

\begin{figure}[H]
    \centering
    \includegraphics[width=\linewidth]{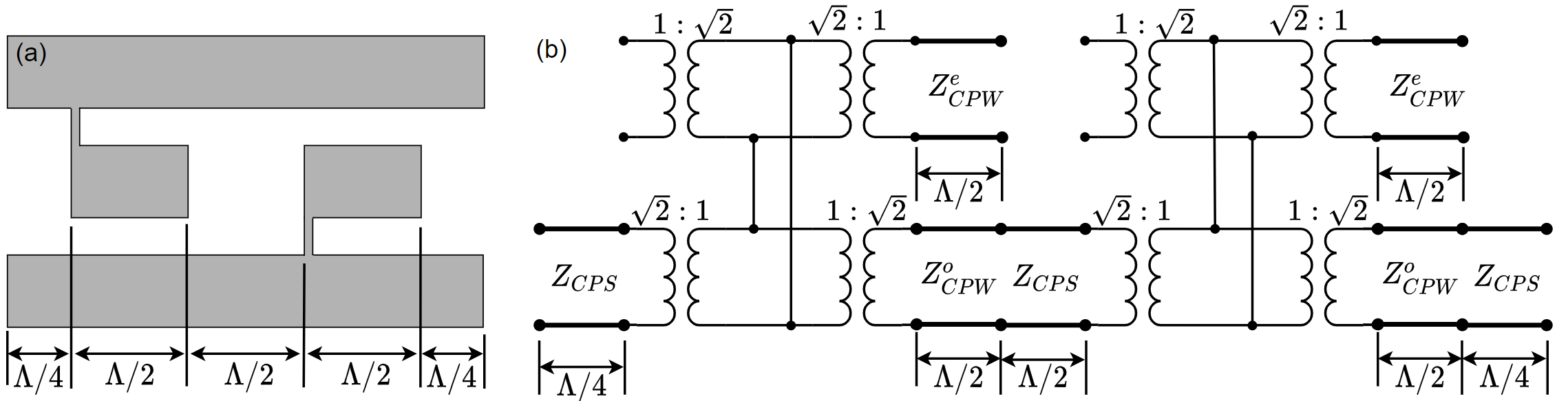}
    \caption{(a) Illustration of the band pass filter conversion which places short circuits between between the conductors of the CPW sections. (b) Circuit representation of the band-pass filter.}
    \label{fig:CPWe_bandpass}
\end{figure}

Figure \ref{fig:CPWe_bandpass}a illustrates the unit cell modification that converts the bandstop filter to a bandpass filter by exciting the even-mode. The short circuit alternates between $Z_A$ and $Z_B$ to ensure that the overall filter remains balanced. Figure \ref{fig:CPWe_bandpass}b is the circuit representation of Fig. \ref{fig:CPWe_bandpass}a.

\begin{figure}[H]
    \centering
    \includegraphics[width=\linewidth]{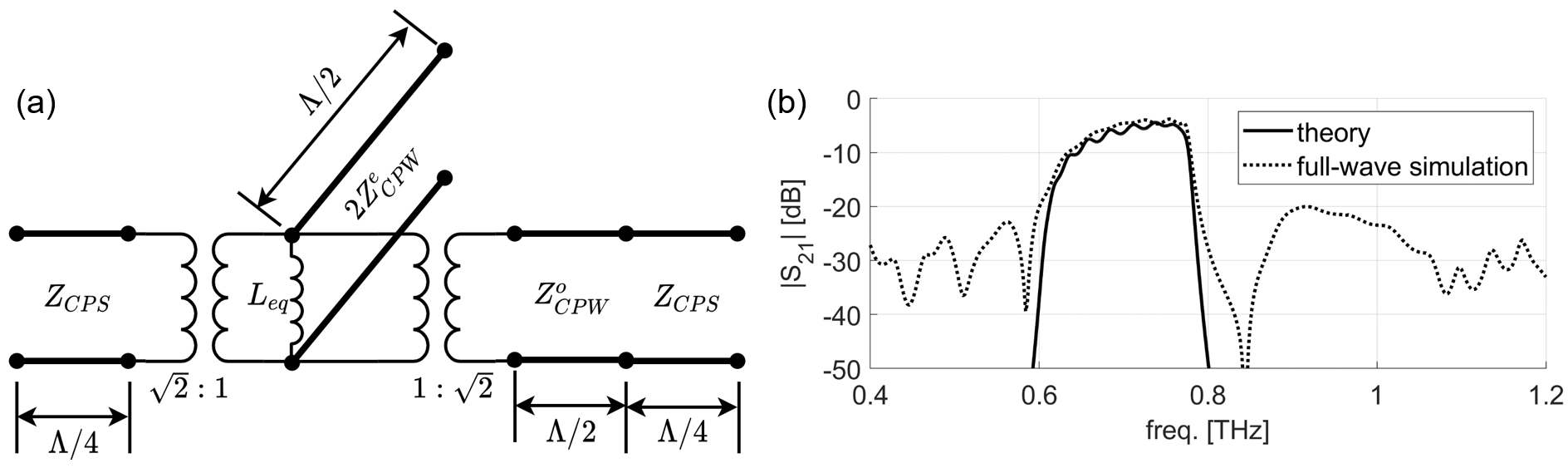}
    \caption{(a) Equivalent circuit of Fig. \ref{fig:CPWe_bandpass}b. (b) Theortical and simulated $|S_{21}|^2$.}
    \label{fig:asymmetric_filter}
\end{figure}

The circuit of Fig. \ref{fig:CPWe_bandpass}b can be simplified to the equivalent circuit of Fig. \ref{fig:asymmetric_filter}a. Using standard methods \cite{pozar2021microwave}, Fig. \ref{fig:asymmetric_filter}a is represented with cascaded ABCD matrices and converted to S-parameters to predict the transmission response. From simulation, we found $L_{eq}$ = 7 pH. We cascaded ten unit cells and compared full-wave simulation and theory in Fig. \ref{fig:asymmetric_filter}b where band-pass behavior is observed. This example demonstrates that exploitation of the multimode characteristics result in unique device possibilities that can be directly integrated into the filter. Lastly, it is possible to use active elements for $Z_A$ and $Z_B$ which would enable active control of the filter performance.

\end{appendices}


\bibliography{main}

\end{document}